\documentclass[12pt]{article}

\usepackage[tbtags]{amsmath}
\usepackage{amsmath}
\usepackage{amssymb}
\usepackage{amsthm}

\newcommand{\al}{\alpha}
\newcommand{\be}{\beta}

\newcommand{\de}{\delta}

\newcommand{\schg}{Schr\"odinger group}
\newcommand{\Schr}{Schr\"odinger}
\newcommand{\di}{differential invariant}
\newcommand{\deq}{differential equation}
\newcommand{\ieq}{invariant equation}
\newcommand{\pde}{partial differential equation}

\newcommand{\vf}{vector field}
\newcommand{\gen}[1]{\partial_{#1}}
\newcommand{\pc}{\psi^\ast}
\newcommand{\dil}{x\gen x+y\gen y+2t\gen t}

\newcommand{\curl}[1]{ \{#1\} }
\newcommand{\free}{i\, \psi_t+\Delta \psi}
\newcommand{\rot}{y\gen x-x\gen y}
\newcommand{\gal}{\mathfrak g}
\newcommand{\Al}{\mathcal A}
\newcommand{\sg}{\mathfrak {sg}}
\newcommand{\heis}{\mathfrak h}
\newcommand{\sch}{\mathfrak {sch}}
\newcommand{\norm}[1]{\vert#1\vert}
\newcommand{\R}{\mathbb{R}}
\newcommand{\C}{\mathbb{C}}
\newcommand\one{\hbox{{1}\kern-.25em\hbox{l}}}

\DeclareMathOperator{\Gr}{G}
\DeclareMathOperator{\Sl}{sl}
\DeclareMathOperator{\SL}{SL}
\DeclareMathOperator{\pr}{pr}
\DeclareMathOperator{\ort}{O}
\DeclareMathOperator{\Sch}{Sch}
\DeclareMathOperator{\SO}{SO}
\DeclareMathOperator{\Heis}{H}
\DeclareMathOperator{\so}{so}
\DeclareMathOperator{\Mink}{M}

\numberwithin{equation}{section}
\begin{document}
\title{{\hfill\normalsize\sf solv-int/9811015}
\vskip1truecm
Nonlinear Evolution Equations Invariant Under\\
Schr\"odinger Group\\ in three-dimensional Space-time}
\author{F. \textsc{G{\"u}ng{\"o}r}\\
\small
\begin{tabular}{c}
Department of Mathematics, Faculty of Science\\ Istanbul
Technical University\\ 80626, Istanbul, Turkey\\
\texttt{gungorf@itu.edu.tr}
\end{tabular}}
\date{}
\maketitle
\begin{abstract}
A classification of all possible realizations of the
Galilei, Galilei-similitude and  \Schr{} Lie algebras in three dimensional
space-time in terms of vector fields under the
action of the group of local diffeomorphisms of the space
$\R^3\times\C$ is presented. Using this result
a variety of general second order evolution equations invariant under
the corresponding groups  are constructed and  their  physical
significance are discussed.
\end{abstract}
\section{Introduction}
A useful tool for treating nonlinear \deq s is the
symmetry method. This includes, among others, generating a
family of nontrivial exact solutions from a known
(trivial) one, reducing the order for ODEs and the number
of independent variables for PDEs and classifying
equations into equivalence classes and hence deriving the
necessary conditions for apparently different equations to
be transformable amongst each other. This method has been
used to obtain exact particular solutions of a large
number of physically important nonlinear \pde s. On the
other hand, if we know the symmetry group of an equation
we can construct other equations which are more general
and  left invariant by the same group. The great advantage
of this inverse approach  is that symmetry reduction technique  can be
used to find some exact solutions of a large class of equations which are
by construction invariant.

In spite of the abundance of work pursuing the direct method
(determining symmetry group and then finding particular solutions of the
reduced equations), there appear only  a few papers devoted to the
alternative problem of constructing the most general equations invariant
under a given symmetry group \cite{rideau1, gungor2, rideau2, heredero,
fush1}. Surprisingly, the complete classification of invariant equations
for many of the groups of physical importance with arbitrary spatial
dimensions is unknown. Though in \cite{fush1} the most general second
order invariant equations for the rotation group $\ort (n)$ as well as
the Euclidean, Poincar\`e and conformal groups are constructed, this
approach seems cumbersome and less systematic. We mention that higher
dimensional invariant equations for the above groups is also unknown. A
more systematic way of constructing general equations invariant under a
group is adopted in \cite{rideau1, rideau2}. An extension of the results
of \cite{rideau1} to a higher dimensional case was carried out in
\cite{gungor2}.

Recently, there has been an explosion of papers devoted to a search for
modified versions of quantum mechanics in which the superposition
principle is no longer valid and the evolution equation is no longer
linear. For instance, Sabatier \cite{sabatier} introduced a set of nonlinear
equations  of the \Schr{} type. In particular, he studied two  classes of
modular equations depending on a real parameter.
Later, Auberson and et al \cite{auberson}
showed that these equations can be linearized by scaling variables. It
turned out that only a subclass of this set, i.e.  the so called modular
class for the free particle case shares the property of being norm conserving
 and being homogeneous,
time reversal invariant with the linear \Schr{} equation. Another remarkable
fact is that, besides being Galilei invariant as mentioned in
\cite{auberson},  they have the same   invariance   group
as the linear \Schr{} equation. In the meanwhile, Doebner and
Goldin \cite{doebner} proposed an eight parameter family of homogeneous nonlinear
\Schr{} type equations on $\R^3,$ including a diffusion term, called
DG-equations from fundamental considerations of local symmetry in
quantum mechanics. They possess nice symmetry properties;
for certain choices of the five parameters they are invariant under a
central extension of the Galilei group. A study of symmetries of the
free DG-equations in terms of gauge invariants and the integrability of
certain sub-families associated with their symmetries has been reviewed in
\cite{nattermann1}. As a matter of fact DG-equations include numerous
modifications of the linear \Schr{} equation. For example, equations studied in
\cite{sabatier} are special cases of DG-equations. One observation that occurs
in  equations proposed as a possible  modification of
the \Schr{} equation from physical considerations only is that they generally
do not obey the Galilean
principle. On the other hand,  Galilei invariant equations are
usually consistent with dilation and conformal invariance. That is why a
group-theoretical justification for this  modification is of vital
importance. From this point of view the construction of all nonlinear
evolution type equations preserving \Schr{} invariance will play an
essential role in modifying \Schr{} equation. Let us emphasize that on physical
grounds the property
of homogeneity is an important ingredient in nonlinear \Schr{}
equations. Indeed, Weinberg  \cite{weinberg} proposed tests of quantum
mechanics by modifying the \Schr{} equation by nonlinear terms and then
imposing a complex homogeneity condition. One can formulate  such a
theory by requiring that the fundamental equations be invariant under a
group isomorphic to the symmetry group of the linear \Schr{} equation,
but omitting the superposition principle.

It is the objective of this paper to construct the
nonlinear second order evolution equations
\begin{equation}\label{1.1}
\psi_t+F(x,y,t,\psi,\pc,\psi_i,\pc_i,\psi_{ij},\pc_{ij})=0,\quad
i,j\in\curl{x,y}.
\end{equation}
invariant under the Galilei, Galilei-similitude and \Schr{} groups.
In \eqref{1.1}, $\psi(x,y,t)$ is a complex function, the
star denotes complex conjugation, the subscripts denote
spatial derivatives and $F$ is a complex function of the
indicated variables. In particular, linear \Schr{} equation
for a free particle
\begin{equation}\label{1.2}
i\, \psi_{t}+\Delta\psi=0
\end{equation}
where $\Delta$ is the Laplace operator in two dimensional Euclidean
space, is contained in the set of equations \eqref{1.1}. It is easy to
show that \eqref{1.2} admits nine dimensional \schg{} and an infinite
dimensional invariant subgroup reflecting the linearity of the equation
(linear superposition principle). An additional symmetry corresponds to
the homogeneity property stating that if $\psi$ is a solution to
\eqref{1.2} then so is $\al \psi$ where $\al$ is a complex constant. A
similar problem for $\Sch(1)$ is solved in \cite{rideau2} and a new
realization of $\Sch(1)$ completing the results of \cite{rideau2} is
given in \cite{zhdanov}. Fushchich et al \cite{fush3} constructed systems
of $(n+1)$-dimensional quasilinear second order evolution equations
invariant under $\sch(n)$. They first make an ansatz about the form of
the equation as a reasonable generalization of a linear equation
and then impose invariance.

The organization of the paper is as follows. In section 2, as a first
step towards classifying \Schr{}  invariant equations, we  classify  all
possible inequivalent realizations of the Galilei $\gal(2,1)$,
Galilei-similitude $\sg(2,1)$ and \Schr{} $\sch(2)$ Lie algebras in
terms of \vf s under the action of local diffeomorphisms of the space
$\R^3\times\C$. In section 3, we obtain second order \di s of the
realized \vf s and hence second order $\sch(2)$ invariant  \deq s. We
also identify $\sch(2)$ invariant equations satisfying homogeneity
condition. We conclude the paper with some remarks and a summary of the
results obtained. Finally, let us emphasize that throughout the paper
invariance under a symmetry group will be in the sense that the equation
is annihilated by second order prolongations of the vector fields on the
solution set. Certainly, this invariance requirement is more restricted
than the case imposed for relativistic equations \cite{rideau1,
gungor2}.  In the present article we shall not take into account the
infinite dimensional Lie group reflecting superposition principle of all
linear \deq s.

\section{Realizations of the Lie algebras by \vf s}
\subsection{\schg{} $\Sch(2)$ and its Lie algebra}
 The \schg{} $\Sch(2)$ is nine parameters local group of transformations
 of the  space $\R^3\times\C$. It is a Lie group isomorphic to
\begin{equation}\label{2.1}
H_2\vartriangleright\curl{\SL(2,\R)\otimes \SO(2)}
\end{equation}
where $\vartriangleright$ denotes a semi-direct product
and $\SL(2,\R)$, $\SO(2)$ and  $\Heis_2$  are the special
linear group, rotation group in the plane and Heisenberg
group, respectively. It is also
subgroup of the conformal group $\ort(4,2)$ of a (2+1)-dimensional
Minkowski space $\Mink(2,1)$ which leaves the Hamilton-Jacobi equation in
two space dimensions invariant \cite{gungor2}. \schg{} appears to be
symmetry group of a variety of physically significant equations. Indeed,
we already mentioned that \eqref{1.2} is invariant under $\Sch(2)$. It is
isomorphic to the symmetry group of two-dimensional heat equation. It is
also symmetry group of the 2-dimensional Navier-Stokes equations in the
case when a linear homogeneous transformation law is imposed on the
pressure.

A recent study of finite dimensional nonrelativistic
conformal groups is given in \cite{negro}.

The corresponding Lie algebra has a basis spanned by two space
translations $P_1, P_2$, two Galilei boosts $B_1,B_2$,
translation of the phase  $M$, time translation $T$,
dilation $D$, rotation $J$ and conformal transformations
$C$. This means that we are dealing with a nine
dimensional Lie algebra $\sch(2)$ which
can be written as a semi-direct sum reflecting the Levi decomposition
\begin{equation}\label{2.2}
\begin{array}{ll}
\sch(2)&=\heis_2 \square\curl{\Sl(2,\R)\oplus\so(2)} \\[.2cm]
       &\sim\curl{P_1,P_2,B_1,B_2,M}\square\curl{T,C,D,J}
\end{array}
\end{equation}
with nonzero commutation relations
\begin{equation}\label{2.3}
\begin{array}{lll}
[P_1,B_1]=M/2,\quad  &  [P_2,B_2]=M/2,&\\[2mm]
[J,B_2]=-B_1,\quad & [J,P_2]=-P_1, &\\[2mm]
[J,B_1]=B_2,\quad & [J,P_1]=P_2, &\\[2mm]
[T,B_j]=P_j,\quad & [D,B_j]=B_j, \quad j=1,2&\\[2mm]
[D,P_j]=-P_j,\quad & [C,P_j]=-B_j,  \quad j=1,2&\\[2mm]
[T,D]=2T,  \quad   & [T,C]=D, & [D,C]=2C.
\end{array}
\end{equation}
In \eqref{2.2}, $\heis_2\sim\curl{P_1,P_2,B_1,B_2,M}$ is the
Heisenberg algebra  with center $M$. The
subalgebra $$\curl{P_1,P_2,T,B_1,B_2,J,M}$$ corresponds to
the extended Galilei algebra, and subalgebra
$$\curl{D,P_1,P_2,T,B_1,B_2,J,M}$$ to the extended
Galilei-similitude algebra.
Let us remark that actually different Levi  decompositions of $\sch(2)$
exist \cite{burdet1}.

The action of the symmetry group on the space
$(x,y,t,\psi,\pc)$ is obtained by integrating the vector
fields of \eqref{2.2} which is referred to as exponentiation of the \vf s.

\subsection{Realizations of Galilei, Galilei-similitude and \Schr{} algebras by
vector Fields}
We classify realizations of $\gal(2,1)$, $\sg(2,1)$ and $\sch(2)$ in terms of
vector fields of the form \begin{equation}\label{2.4} {\mathbf v}=\xi \gen
x+\eta \gen y+\tau \gen t+\sigma \gen \psi+\sigma^*\gen {\psi^*} \end{equation}
where $\xi,\eta,\tau,\sigma,\sigma^*$ are functions of $x,y,t,\psi,\psi^*$ with
$\xi,\eta,\tau\in\R$, and $\sigma,\sigma^*\in\C$, under local diffeomorphisms
of the space $\R^3\times\C$  parametrized by the variables $(x,y,t,R,\phi)$
with $x,y,t$ space-time coordinates and $\psi=R e^{i\phi}$ the wave function.
Two realizations will be equivalent  if the corresponding vector fields can be 
transformed into each other by arbitrary smooth invertible changes of the 
independent and dependent variables:
\begin{equation}
\begin{array}{l}
\tilde x=X(x,y,t,R,\phi),\quad \tilde y=Y(x,y,t,R,\phi),
\quad \tilde t=T(x,y,t,R,\phi)\\
\tilde R=\Psi(x,y,t,R,\phi),\quad \tilde\phi=\Phi(x,y,t,R,\phi).
\end{array}
\end{equation}

\subsubsection{The extended Galilei algebra}
We start from the four-dimensional abelian algebra
$\curl{P_1,P_2,T,M}$. It is immediate to rectify these vector fields (see
\cite{olver1, olver2} for the vector field rectification theorem) up to
the diffeomorphisms as
\begin{equation}\label{2.5}
\Al_4 : \curl{P_1=\gen
x,P_2=\gen y, T=\gen t, M=\gen \phi}.
\end{equation} $\Al_4$  generates
space-time translations and translation of phase of the wave function.
Obviously, these vector fields remain invariant under
\begin{subequations}\label{2.6}
\begin{gather}
\tilde x=x+g(R), \quad \tilde y=y+h(R), \quad \tilde
t=t+f(R),\\[.2cm]
\tilde \phi=\phi+\Phi(R), \quad \tilde R=\rho(R)
\end{gather}
\end{subequations}
where $f,g,h,\Phi$ and $\rho$ are arbitrary functions of $R$. 
Let us mention that the standard vector-field realizations in 2+1 dimensional
space-time can be found in \cite{barut}. If the Galilei
boost $B_1$ having the form \eqref{2.4} is subjected to the commutation
relations involving $B_1$ and  elements of  the already realized algebra
$\Al_4$ and is further simplified by \eqref{2.6}, one finds precisely two
inequivalent types of realizations of $B_1$:
\begin{subequations}\label{2.7}
\begin{align}
B_1^{(1)}&=t\gen x+f_2(R)\gen y+f_3(R)\gen t+x/2\gen \phi\\
B_1^{(2)}&=t\gen x+R\gen R+x/2\gen \phi.
\end{align}
\end{subequations}
Next we realize $B_2$  written in the form \eqref{2.4}. Imposing commutation
relations and simplifying by transformations leaving the algebra $\curl{\Al_4,
B_1^{(j)},\quad j=1,2}$ invariant it follows that $B_1$ can be further
extended and we have
\begin{subequations}\label{2.8}
\begin{align}
B_1^{(1)}&=t\gen x+f_2(R)\gen y+x/2\gen \phi\label{2.8a}\\
B_2^{(1)}&=f_2(R)\gen x+(t+g_2(R))\gen y+y/2\gen \phi \label{2.8b}
\end{align}
and
\begin{align}
B_1^{(2)}&=t\gen x+R\gen R+x/2\gen \phi \label{2.8c}\\
B_2^{(2)}&=t\gen y+\mu R\gen R+y/2\gen \phi  \label{2.8d}
\end{align}
\end{subequations}
where $\mu$ is a constant. Proceeding as above we see that the extension
of $\curl{\Al_4, B_1^{(2)},B_2^{(2)}}$ to rotations is not possible.
However, for \eqref{2.8a}-\eqref{2.8b} we find
\begin{equation}\label{2.9}
J=y\gen
x-x\gen y+j_3(R)\gen t+j_4(R)\gen R+j_5(R)\gen \phi
\end{equation}
with $f_2$ and $g_2$ related by $$4f_2f_2'+g_2g_2'=0$$ and $j_3=2f_2$,
$j_4, j_5$ arbitrary. Consequently, we obtain a single realization of the
Galilei algebra with infinitesimal generators \eqref{2.5},
\eqref{2.8a}-\eqref{2.8b} and \eqref{2.9} depending on three arbitrary
functions.

\subsubsection{The extended Galilei-similitude algebra} Let us
now add a dilation generator $D$ of the form \eqref{2.4}. The commutation
relations between $\curl{P_1,P_2,B_1,B_2,T,M,D}$ restricts $D$ to
\begin{equation}\label{2.10}
D=\dil+d_4(R)\gen R+d_5(R)\gen \phi.
\end{equation}
Commuting $J$ with $D$ yields
$$j_4(R)=a d_4(R) \quad \text{and} \quad j_5(R)=a d_5(R), \qquad
a=\text{const.}$$ and forces $B_1^{(1)}$, $B_2^{(1)}$ and $J$ to be
\begin{gather}
B_1^{(1)}=t\gen x+x/2\gen\phi,\quad B_2^{(1)}=t\gen
y+y/2\gen\phi,\nonumber\\[.3cm]
J=y\gen x-x\gen y+a(d_4(R)\gen
R+d_5(R)\gen\phi).\nonumber
\end{gather}
The form of $D$ is further restricted by transformations
leaving the algebra $$\curl{P_1,P_2,B_1,B_2,M,T,J}$$ invariant.
Finally, we obtain two inequivalent realizations of the
Galilei-similitude algebra which depend on one arbitrary
function
\begin{subequations}\label{2.11}
\begin{align}
sg^{1}(d_0)&:
\begin{cases}
B_1&=t\gen x+x/2 \gen \phi,\quad B_2=t\gen y+y/2\gen\phi\\
          J&=y\gen x-x\gen y-a R/2 \gen R+a d_0(R)\gen \phi\\
          D&=\dil-R/2\gen R+d_0(R)\gen \phi
\end{cases}\label{2.11a}\\[.5cm]
sg^{2}(\de_0)&:
\begin{cases}
B_1&=t\gen x+x/2 \gen \phi,\quad B_2=t\gen y+y/2\gen\phi\\
J&=y\gen x-x\gen y+a\de_0(R)\gen \phi\\
D&=\dil+\de_0(R)\gen\phi.
\end{cases}\label{2.11b}
\end{align}
\end{subequations}
\subsubsection{The  \Schr{} algebra}
The above obtained realizations of $\sg(2,1)$ can be
further extended to $\sch(2)$ by adding the nonrelativistic conformal
generator $C$. We start from \eqref{2.11} and repeat the routine steps of
imposing commutation relations and simplifying by transformations leaving
the $\sg(2,1)$ algebra realizations unchanged. Omitting  the lengthy
details we present the final results only:
\begin{subequations}\label{2.12}
\begin{align}
sch^1(f;\al,\be)&:
\begin{cases}
J&=y\gen x-x\gen y\\
D&=\dil-\displaystyle\frac{R}{2}\gen
R+f(R)\gen \phi\\
C&=xt\gen x+yt\gen y+t^2\gen
t+\Bigl(-\displaystyle\frac{tR}{2}+\al R^{-3}\Bigr)\gen
R\\&\quad+\Bigl((x^2+y^2)/4+t f(R)+R^{-4}(\be-2\al f(R))\Bigr)\gen \phi
\end{cases}\label{2.12a}\\[.5cm]
sch^2(g)&:
\begin{cases}
J&=y\gen x-x\gen y-\displaystyle\frac{a R}{2}\gen R+a g(R)\gen \phi\\
D&=\dil-\displaystyle\frac{R}{2}\gen R+g(R)\gen \phi\\
C&=xt\gen x+yt\gen y+t^2\gen t-\displaystyle\frac{tR}{2}\gen R
+\Bigl((x^2+y^2)/4+tg(R)\Bigr)\gen \phi
\end{cases}\label{2.12b}\\[.5cm]
sch^3(h)&:
\begin{cases}
J&=y\gen x-x\gen y+a h(R)\gen \phi\\
D&=\dil+h(R)\gen\phi\\
C&=xt\gen x+yt\gen y+t^2\gen t+\Bigl((x^2+y^2)/4+t h(R)\Bigr)\gen\phi.
\end{cases}\label{2.12c}
\end{align}
\end{subequations}
We observe that all of the above realizations characterize a class of
algebras corresponding to an arbitrary function and constants and
generate fiber-preserving transformations. Using the relations
$$R\gen R=\psi\gen \psi+\pc\gen \pc,\quad
\gen\phi=i(\psi\gen\psi-\pc\gen\pc)$$ we can express all the generators
involving $R$ and $\phi$ (modulus and phase of the wave) in terms of the
original variables $x,y,t,\psi,\pc$.
\section{Differential invariants and invariant equations}
In this section we obtain second order differential invariants 
under the particular cases of the realizations obtained in 
section 2.2 and hence  invariant
differential equations admitting \Schr{} symmetry.

A brief review of the \di s and  fundamental theorem that is essential
for constructing  \ieq s was outlined in \cite{gungor2}. For a full
discussion of the relevant definitions and theorems with proofs we refer
to the contemporary literature \cite{olver1,olver2,ovsiannikov,bluman}.
In order to
make the present paper self-contained we recall some definitions which
will be needed in the derivation of \di s. Let $\Gr$ be a local Lie group
of transformations acting on the space of independent and dependent
variables $X\otimes U$ and  $\Gr^{(n)}=\pr^{(n)}\Gr$ denote the prolonged
group action on the jet space $J^n$ whose coordinates are denoted by
$(x,u^{(n)})$. The space of infinitesimal generators of $\Gr$, i.e. its
Lie algebra will be denoted by $\gal$ with associated prolongation
$\gal^{(n)}=\pr ^{(n)} \gal$. Recall that an absolute \di{} of order
$r\le n$ is a scalar function $I: J^n\to \R$ which satisfies
\begin{equation}\label{3.1} I(g^{(n)}.(x,u^{(n)}))=I(x,u^{(n)})
\end{equation}
for all
$g^{(n)}\in \Gr^{(n)}$ and all $(x,u^{(n)})\in J^n$. Since any function
$F(I_1,I_2,\ldots\, I_r)$ of a collection of \di s $\curl{I_1,I_2,\cdots,I_r}$
is also a \di{}, we classify \di s up to functional independence. A complete
set of functionally independent \di s will be called fundamental invariants.
Once we have found such a complete set, any other \di{} can be expressed as a
function of these invariants. Let  $\mathbf v$ be a one-parameter group of
transformations acting on $X\otimes U$, the associated $n$-th order
prolonged vector field  $\pr^{(n)} \mathbf v$ is the vector field on the
jet space $J^n$. The infinitesimal version of \eqref{3.1} can be written as
\begin{equation}\label{pr}
\pr ^{(n)} \mathbf v(I)=0
\end{equation}
for every prolonged vector field $\pr^{(n)}\mathbf v$.
This implies that
if $\curl{\mathbf v_i},\;i=1,2,\ldots,r$ form a basis for the symmetry algebra
then  the \di s  are found by
solving an overdetermined  system of homogeneous, first order linear \pde s
of \eqref{pr} with $\mathbf v$ replaced by the basis elements $\mathbf v_i$.
A solution to this system is a differential function depending on $n$-th order
jet variables.
One way to proceed towards finding \di s is to first solve  the system for
$\mathbf v_1$ which
amounts to obtaining the \di s of $\mathbf v_1$ and next to re-express
the remaining vector fields in terms of these invariants as coordinates and
to find invariants of $r-1$ vector fields, namely to solve inductively the
rest of the system containing $r-1$ equations using the same procedure.
The general formula for the $n$-th
prolongation of a \vf{} is given, e.g. in \cite{olver1}.
Since we concentrate on second order
\deq s here, we need expressions for second prolongation
$\pr^{(2)}\mathbf v_i$ for each infinitesimal generator $\mathbf v_i$.
The most drudgery part of the present paper, i.e., the calculation of
the explicit expressions for $\pr^{(2)}\mathbf v_i$
has been eliminated  using MATHEMATICA.

The  general form of an invariant evolution equation will be
$$\psi_t+F(I_1,\cdots,I_k)=0$$ where $\curl{I_1,\cdots,I_k}$ are
fundamental invariants.  The function $F$ will be obtained as a solution
to the \pde s
$$\pr^{(2)}\mathbf v_i(\psi_t+F)=0,\quad \text{whenever}\quad \psi_t=-F$$
for every vector field $\mathbf v_i$ in the basis and $F$ has a priori the
form in \eqref{1.1}.
In other words, each element chosen from the  realized algebra provides a
first order
linear \pde{} for the function $F$. Solving this overdetermined system
we express $F$ in
terms of fundamental invariants whenever possible and hence construct
\ieq s. In the following we apply this scheme to obtain the Galilei,
Galilei-similitude and \Schr{} invariant equations, respectively.
\subsection{The extended Galilei invariant equations}
Consider the standard realization
\begin{equation} \begin{split} \curl{\curl{\Al_4}, B_1&=t\gen x+\frac{i
x}{2}(\psi\gen\psi-\pc\gen\pc), B_2=t\gen y+\frac{i
y}{2}(\psi\gen\psi-\pc\gen\pc),\\J&=\rot}. \end{split} \end{equation}
Invariance under $\Al_4$ will restrict the form of \eqref{1.1} to
\begin{equation}\label{3.2}
\psi_t+\psi_{xx}+F(I_j)\psi=0,\qquad j=1,11
\end{equation}
\begin{align}
I_1&=\psi\pc=\norm{\psi}^2,  & &\nonumber\\
I_2&=\pc\psi_x,& I_3&=\psi\pc_x=I_2^*\nonumber\\
I_4&=\pc\psi_{xx},& I_5&=\psi\pc_{xx}=I_4^*\nonumber\\
I_6&=\pc\psi_y,& I_7&=\psi\pc_y=I_6^*\nonumber\\
I_8&=\pc\psi_{yy},& I_9&=\psi\pc_{yy}=I_8^*\nonumber\\
I_{10}&=\pc\psi_{xy},& I_{11}&=\psi\pc_{xy}=I_{10}^*\nonumber.
\end{align}
A further requirement of invariance under $B_1$ reduces eleven invariants to
ten and \eqref{3.2} to
\begin{equation}\label{3.3}
i\, \psi_t+\psi_{xx}+F(J_k)\psi=0,\qquad k=1,10
\end{equation}
\begin{align}
J_1&=I_1,& J_2&=I_2+I_3=(\norm\psi^2)_x\nonumber\\
J_3&=I_3^2-I_1I_5,& J_4&=I_2^2-I_1I_4=J_3^*\nonumber\\
J_5&=I_1I_{10}+I_6I_3, &J_\mu&=I_\mu,\quad \mu=6,7,8,9\nonumber\\
J_{10}&=I_1I_{11}+I_6I_3=J_5^*. && \nonumber
\end{align}
Imposing invariance under $B_2$ will reduce ten invariants to nine and
\eqref{3.3} to
\begin{equation}\label{3.4}
\free+F(K_\lambda)\psi=0,\qquad \lambda=1,9
\end{equation}
\begin{align}
K_\sigma&=J_\sigma, \quad \sigma=1,2,3,4 &&\nonumber\\
K_5&=J_5+J_{10}, & K_6&=J_6+J_7=(\norm\psi^2)_y\nonumber\\
K_7&=J_6^2-J_1J_9, & K_8&=J_7^2-J_1J_9=K_7^*\nonumber\\
K_9&=J_5-J_2J_6.&&\nonumber
\end{align}
where $K_1$, $K_2$, $K_5$ and $K_6$ are real.
Finally if we require \eqref{3.4} be rotationally invariant we are led to
the following extended Galilei-invariant equation
\begin{equation}\label{3.5}
\free+F(L_1,L_2,L_3,L_4,L_5)\psi=0
\end{equation}
\begin{align}
L_1&=K_1, & L_2&=K_2^2+K_6^2\nonumber \\
L_3&=K_3+K_8, & L_4&=K_4+K_7=L_3^*\nonumber\\
L_5&=K_9^2-K_4K_7\nonumber.
\end{align}
where $L_1$ and $L_2$ are real.
In terms of the original variables we have
\begin{align}
L_1&=\norm\psi^2\nonumber\\
L_2&=\curl{(\norm\psi^2)_x}^2+\curl{(\norm\psi^2)_y}^2=(\nabla\norm\psi^2)^2
=4\norm\psi^2{(\nabla\norm\psi)}^2\nonumber\\
L_3&=\psi^2({\pc_x}^2+{\pc_y}^2)-\norm\psi^2\psi\Delta\pc=
\psi^2{(\nabla\pc)}^2-\norm\psi^2\psi\Delta\pc\nonumber\\
L_4&=L_3^*={\pc}^2({\psi^2_x}+{\psi^2_y})-\norm\psi^2\pc\Delta\psi
={\pc}^2{(\nabla\psi)}^2-\norm\psi^2\pc\Delta\psi\nonumber\\
L_5&=\curl{\norm\psi^2(\pc\psi_{xy}+\psi_y\pc_x)-\pc\psi_y
{\norm\psi^2_x}}^2
-{\pc}^4({\psi^2_x}-\psi\psi_{xx})({\psi^2_y}-\psi\psi_{yy})\nonumber
\end{align}
In particular, \eqref{3.5} includes a physically relevant
subclass of nonlinear equations of the form
$$\free+f(\norm\psi)\psi=0.$$

\subsection{The extended Galilei-similitude invariant equations}
We now add the requirement that \eqref{3.5} be invariant under the dilation
generator
$$D_1=\dil-2/k(\psi\gen\psi+\pc\gen\pc)\quad k\ne 0$$
which is equivalent to $D$ of \eqref{2.12a} by a simple change of variable
with $f=0$ and obtain the \ieq{}
\begin{subequations}
\begin{gather}\label{3.6}
\free+L_1^{k/2} F(R_1,R_2,R_3)\psi=0  \\[.5cm]
R_1=\frac{L_3}{L_1^{(k+4)/2}},\quad
R_2=\frac{L_2}{L_1^{(k+4)/2}},\quad R_3=\frac{L_4}{L_1^{(k+4)/2}}= R_1^*
\end{gather}
\end{subequations}
where $F$ is an arbitrary complex
function. In particular, setting $F=-\lambda=(\text{const.})$ we obtain the
nonlinear \Schr{} equation with power nonlinearity
\begin{equation}\label{3.7}
\free=\lambda \norm\psi^k\psi.
\end{equation}
We see that the cubic and quintic \Schr{} equations
will belong to the above class of Galilei-similitude \ieq s.
These type of nonlinear generalizations arise in applications as diverse as nonlinear
optics, wave propagation in water, interactions of laser beams with
plasma, turbulence and many others.

When we
extend the Galilei invariant equation by $$D=\dil$$ we get the \ieq{}
corresponding to the realization $sg^2(0)$ of \eqref{2.11b}:
\begin{subequations}\label{3.10}
\begin{gather}
\free +L_2 F(S_1,S_2,S_3)\psi=0\\[.5cm]
S_1=L_1,\quad S_2=\frac{L_3}{L_2},\quad S_3=
\frac{L_4}{L_2}=S_2^*.
\end{gather}
\end{subequations}
\subsection{The \Schr{} invariant equations} To construct
conformally invariant evolution equations we should add the requirement of
conformal invariance generated by $C$.  This will further restrict the
form of the arbitrary functions occurring in the Galilei-similitude
invariant equations. Indeed, for  the standard realization $sch^1(0;0,0)$
the requirement that Eq. \eqref{3.6} be invariant under
$$C=x t\gen x+y t\gen y+t^2\gen t+\curl{
-2/k t+i(x^2+y^2)/4}\psi\gen\psi-\curl{2/k t+i(x^2+y^2)/4}\pc\gen\pc$$
leads, after some lengthy calculations, to the \ieq{}
\begin{subequations}\label{3.11}
\begin{gather}
\free+L_1^{k/2}\Bigl[(1-\frac{2}{k}){R_3}+H(\Sigma _1,\Sigma _2)\Bigr]\psi=0\\[.5cm]
\Sigma _1=R_2=L_1^{-(k+4)/2}L_2,\quad \Sigma
_2=R_1+R_3=L_1^{-(k+4)/2}(L_3+L_4).
\end{gather}
\end{subequations}
where $H$ is an arbitrary complex function of two real variables.
For $k=2$  equation \eqref{3.11}  reduces to
\begin{subequations}\label{3.12}
\begin{gather}
\free+L_1 H(\Sigma _1,\Sigma _2)\psi=0\\[.5cm]
\Sigma _1=L_1^{-3}L_2,\qquad \Sigma _2=L_1^{-3}(L_3+L_4).
\end{gather}
\end{subequations}
Eq. \eqref{3.12} includes particularly the cubic \Schr{} equation
\begin{equation}\label{3.13}
\free=\mu \norm\psi^2\psi.
\end{equation}
Using the identity
$$L_2-(L_3+L_4)=\norm\psi^2\Delta\norm\psi^2$$
yields
$$\Sigma_2=\Sigma_1-\frac{\Delta\norm\psi^2}{\norm\psi^4}$$
and so Eq. \eqref{3.12} can be expressed as
\begin{equation}\label{3.14}
\free+L_1 H_1(\Omega_1,\Omega_2)\psi=0
\end{equation}
where $H_1$ is arbitrary and
$$\Omega_1=\frac{(\nabla\norm\psi)^2}{\norm\psi^4},\qquad \Omega_2=
\frac{\Delta\norm\psi^2}{\norm\psi^4}.$$
In particular, when $H_1$ is restricted to be linear in $\Omega_1, \Omega_2$,
namely
$$H_1=A\Omega_1+B \Omega_2+C,\qquad A,B,C\in\C$$
then we obtain
$$\free+\norm\psi^{-2}\Bigl(A(\nabla\norm\psi)^2+B\Delta\norm\psi^2
+C\norm\psi^4\Bigr)\psi=0$$
containing the following $\sch(2)$ invariant equations
\begin{subequations}\label{3.15}
\begin{equation}\label{mod1}
\free=A\; \frac{(\nabla\norm\psi)^2}{\norm\psi^2}\psi,
\end{equation}
\begin{equation}\label{mod2}
\free=B\; \frac{\Delta\norm\psi^2}{\norm\psi^2}\psi
\end{equation}
and the cubic \Schr{} equation.
It is clear that
\begin{equation}\label{mod3}
\free=B\; \frac{\Delta\norm\psi}{\norm\psi}\psi.
\end{equation}
belongs to the class \eqref{3.14}.
The nonlinear terms on the right side describe dissipative and diffusion
processes in quantum mechanics.
Also, setting $H_1=A(\Omega_1-\Omega_2)$ and using 
$\Delta\ln \norm\psi=\Omega_2-\Omega_1$ we obtain
\begin{equation}\label{log}
\free=A (\Delta\ln\norm\psi)\psi.
\end{equation}
\end{subequations}
The modular class of nonlinear equations of \eqref{3.14} are time reversal 
invariant,
namely invariant under the transformations $t\to -t,\quad \psi\to \pc$.
Eq. \eqref{mod3} was proposed as a stochastic interpretation of quantum
mechanical vacuum dissipative effects \cite{vigier}. Also, equations
\eqref{mod3}-\eqref{log} fall into the class of equations introduced by Sabatier
\cite{sabatier} and DG-equations \cite{doebner}. In particular, these equations
were shown to be linearizable for any $A\in \R$ \cite{auberson}.
In passing, let us mention that
for $A>1$ logarithmic \Schr{} equation \eqref{log} admits a solitary wave
solution
propagating without deformation. A variety of equations proposed as mathematical
models of quantum theory appear to be special cases of the \Schr{} invariant
nonlinear  equation \eqref{3.14}.

Furthermore, if one specializes the arbitrary function $H_1$ in
\eqref{3.12} to be
$$H=A+B\; \Omega_1^{1/2}$$
then Eq. \eqref{3.12} has the form
\begin{equation}\label{3.16}
\free+[A\norm\psi^2+B({\norm\psi}^2_x+{\norm\psi}^2_y)^{1/2}]\psi=0.
\end{equation}
It is natural to call Eq. \eqref{3.16} two dimensional Eckhaus equation.
One dimensional Eckhaus equation is defined to be
$$i\;\psi_t+\psi_{xx}+(A\: \norm\psi^4+B\: {\norm\psi}^2_x)\psi=0$$
where $A, B$ are arbitrary complex numbers.
For $\norm B^2=4A$ with $A$ real, the above equation was
shown to be linearizable, namely   equivalent to the linear \Schr{}
equation and to have the Painlev\'e property \cite{calogero, clarkson1}.
It should be interesting to
study the integrability and linearizability properties of \eqref{3.16}.

In addition to \Schr{} invariance we can construct subsets
of \eqref{3.12} satisfying the homogeneity condition. When this is
the case we find that $H$ has the form
\begin{equation}
H=\Sigma _1\; h(\frac{\Sigma _1}{\Sigma _2}).
\end{equation}
On the other hand, the Galilei-similitude \ieq{} \eqref{3.10} can not be
extended to conformal transformations. This implies that there is no
equation invariant under the realization \eqref{2.12c}. However, notice
that the arbitrary function in \eqref{3.10} can be specified so that
equation satisfies homogeneity condition, namely
\begin{equation}
\free+\frac{L_2}{L_1^2}\; F(S_2,S_3)\psi=0
\end{equation}
with $S_2, S_3$ as in \eqref{3.10}.
Let us comment that, here
as opposed to the one dimensional case, no quintic \Schr{} equation
invariant under \schg{} is obtained. Quintic \Schr{} equation is typical
for one space dimension.

Since all of the realizations obtained in section 2
generate fiber-preserving transformations, namely the coefficients
multiplying $\gen x$, $\gen y$ and $\gen t$ do not depend on $\psi,\pc$,
in principle one can always obtain  equations invariant under the less
standard groups involving an arbitrary function  of a single real
variable $\psi\pc$ by integrating a system of coupled first order \pde s.
But with this generality \ieq s are too complicated and we have not
attempted to write them out here.
\section{Conclusions}
The results of
this article can be summarized as follows. In section 2 we performed a
classification of all possible realizations of the extended Galilei,
Galilei-similitude and \Schr{} algebras in 2+1 space-time dimensions in terms of
\vf{}
under local diffeomorphisms of $\R^3\times\C$. The realizations obtained
depend on  arbitrary functions of a real variable $\psi\pc$ and
constants and the corresponding group transformations are necessarily
fiber preserving.

In section 3 we constructed  evolution type equations
invariant under the groups corresponding to the  realized algebras with
the arbitrary functions set equal to zero. In other words, we obtained a 
variety
of second order equations of the form \eqref{1.1}  invariant
under the symmetry group of the free \Schr{} equation, omitting the
infinite dimensional Lie group that reflects linear superposition
principle. Thus, we have shown that how general a \Schr{} invariant
equation can be. In particular, we obtained a class of equations which might be
candidates for possible generalizations of quantum mechanics for a scalar
particle in (2+1)-dimensional space-time. Such general equations
need a  careful physical interpretation.

In the present study, as in \cite{rideau2}, we
obtain rather general invariant equations involving an arbitrary function
of three or two variables (invariants). But, contrary to the (1+1)-dimensional
case, we do not obtain quintic \Schr{} equation invariant
under any realization of the \schg{}. Let us mention that some realizations
of \cite{rideau2} do not have counterparts in three dimensional case
and vice versa. One thing that is common to both dimensions is that
all realized algebras involve an arbitrary function of a real variable,
namely they characterize a class of algebras and that they correspond to
fiber preserving transformations. The latter property simply means  that
the transformed independent variables under the action of the
corresponding group will only depend  on the old ones, but not on the
wave function. When the arbitrary function figuring in the most general
\Schr{} invariant equation of the standard realization is restricted to 
certain
subfamilies we recovered several invariant equations which fit in the class
proposed in Ref. \cite{sabatier, auberson, doebner}. 
Another special case is two
dimensional analogue of the integrable Eckhaus equation that was first
introduced in \cite{calogero} and generalized in \cite{clarkson1}.

The knowledge that the equations are by construction
invariant under space-time symmetry groups ensures us that
we can always apply the symmetry reduction method to find
exact particular solutions called group invariant
solutions.
For example, a recent study \cite{sciarrino} has been devoted to
symmetries and solutions of the vector \Schr{} equation
with cubic nonlinearity in two space dimensions.
In order to be able to perform symmetry reduction
systematically we need to know a classification of the
subalgebras of the corresponding symmetry algebra into
conjugacy classes, under the action of the symmetry group.
The subalgebras of \Schr{} algebra $\sch(2)$ were
classified by G. burdet et al. in \cite{burdet1}.
In a separate paper we plan
to investigate a classification of symmetry reductions and
their solutions for a specific class of invariant
equations such as those of the form \eqref{3.15}. Especially, we would
like to return to the integrability properties of \eqref{3.16}.
\section*{Acknowledgement}
The author would like to thank the refrees for their useful comments which 
improved the presentation of the paper.

\begin{thebibliography}{10}

\bibitem{rideau1}
G.~Rideau and P.~Winternitz.
\newblock {\em J. Math. Phys.}, 31(5):1095--1105, 1990.

\bibitem{gungor2}
F.~G{\"u}ng{\"o}r.
\newblock {\em J. Phys. A: Math. Gen.}, 31:697--706, 1998.

\bibitem{rideau2}
G.~Rideau and P.~Winternitz.
\newblock {\em J. Math. Phys.}, 34(3):558--570, 1992.

\bibitem{heredero}
R.H. Heredero and P.J. Olver.
\newblock {\em J. Math. Phys.}, 37(12):6414--6438, 1996.

\bibitem{fush1}
W.I. Fushchich and I.A. Yegorchenko.
\newblock {\em Acta Appl. Math.}, 28:69--92, 1992.

\bibitem{sabatier}
P.C. Sabatier.
\newblock {\em Inverse Problems}, 6:L47--L53, 1990.

\bibitem{auberson}
G.~Auberson and P.C. Sabatier.
\newblock {\em J. Math. Phys.}, 35(8):4028--4040, 1994.

\bibitem{doebner}
H-D. Doebner and G.~A. Goldin.
\newblock {\em J. Phys. A: Math. Gen.}, 27:1771--1780, 1994.

\bibitem{nattermann1}
P.~Nattermann and H-D. Doebner.
\newblock {\em J. Nonlin. Math. Phys.}, 3(3-4):302--310, 1996.

\bibitem{weinberg}
S.~Weinberg.
\newblock {\em Ann. Phys.}, 194:336--386, 1989.

\bibitem{zhdanov}
R.Z. Zhdanov and W.I. Fushchych.
\newblock {\em J. Nonlin. Math. Phys.}, 4(3-4):426--435, 1997.

\bibitem{fush3}
W.I. Fushchich and R.M. Cherniha.
\newblock {\em J. Phys. A: Math. Gen.}, 28:5569--5579, 1995.

\bibitem{negro}
J.~Negro, M.A. del Olmo, and A.~Rodriguez-Marco.
\newblock {\em J. Math. Phys.}, 38(7):3786--3809, 1997.

\bibitem{burdet1}
G.~Burdet, J.~Patera, M.~Perrin, and P.~Winternitz.
\newblock {\em Ann. Sc. Math. Quebec}, II(1):81--108, 1978.

\bibitem{olver1}
P.J. Olver.
\newblock {\em Applications of Lie Groups to Differential Equations}.
\newblock Springer, New York, 1991.

\bibitem{olver2}
P.J. Olver.
\newblock {\em Equivalence, Invariants and Symmetry}.
\newblock Cambridge University Press, Cambridge, 1995.

\bibitem{barut}
A.O. Barut and R. Raczka.
\newblock {\em Theory of Group Representations and Applications}.
\newblock Polish Sci. Publ., Warsaw, 1980.

\bibitem{ovsiannikov}
L.V. Ovsiannikov.
\newblock {\em Group Analysis of Differential Equations}.
\newblock Academic Press, New York, 1982.

\bibitem{bluman}
G.W. Bluman and S.~Kumei.
\newblock {\em Symmetries and Differential Equations}.
\newblock Springer, New York, 1989.

\bibitem{vigier}
J.P. Vigier.
\newblock {\em Phys. Lett. A}, 135:99--105, 1989.

\bibitem{calogero}
F.~Calogero and S.~De Lillo.
\newblock {\em Inverse Problems}, 4:L33--L37, 1988.

\bibitem{clarkson1}
P.A. Clarkson.
\newblock {\em Nonlinearity}, 5:453--472, 1992.

\bibitem{sciarrino}
A.~Sciarrino and P.~Winternitz.
\newblock {\em Il Nuovo Cimento}, 112B(6):853--871, 1997.

\end{thebibliography}

\end{document}